\documentstyle[epsfig]{aipproc}

\begin{document}
\title{Supernova rates in Abell galaxy clusters and implications 
for metallicity}

\author{Avishay Gal-Yam$^{1}$ and Dan Maoz$^{1}$}
\address{$^1$School of Physics and Astronomy and Wise Observatory, 
Tel Aviv University, Tel Aviv 69978, Israel}

\maketitle

\begin{abstract}
Supernovae (SNe) play a critical role in the metal enrichment of the 
intra-cluster medium (ICM) in galaxy clusters. Not only are SNe the main 
source for metals, but they may also supply the energy to eject enriched 
gas from galaxies by winds. However, measurements of SN rates in galaxy 
clusters have not been published to date. We have initiated a program to 
find SNe 
in 163 medium-redshift ($0.06<z<0.2$) Abell clusters, using the Wise 
Observatory 1m telescope. We report here our on first results and describe
our main scientific goals.  
Following the discovery of a SN in the remote periphery (78 Kpc) of the cD
galaxy of Abell 403, we discuss a novel explanation for the centrally 
enhanced metal abundances indicated by X-ray observations of galaxy clusters.
\end{abstract}

\section*{The project and main scientific objectives}

We have used the Wise Observatory 1m telescope to monitor monthly a sample
of 163 rich (richness class $R > 0$) Abell galaxy clusters, with medium
redshift ($0.06<z<0.2$) northern declination $(\delta > 0)$ and small angular 
size $(r < 20')$. We have also observed "blank" flanking fields for a 
sub-sample of the clusters. These will be used to study the cluster vs. field
SN rates, and to estimate the luminosity contributed by the cluster in each 
field. We have used  unfiltered ("clear") observations to achieve maximum
sensitivity, and have a characteristic limiting magnitude of $R\sim22$.
Variable objects are discovered by image subtraction. New subtraction methods
have been developed for use in this project (see fig. 1).\\
 
\begin{figure}[t!] 
\centerline{\epsfig{file=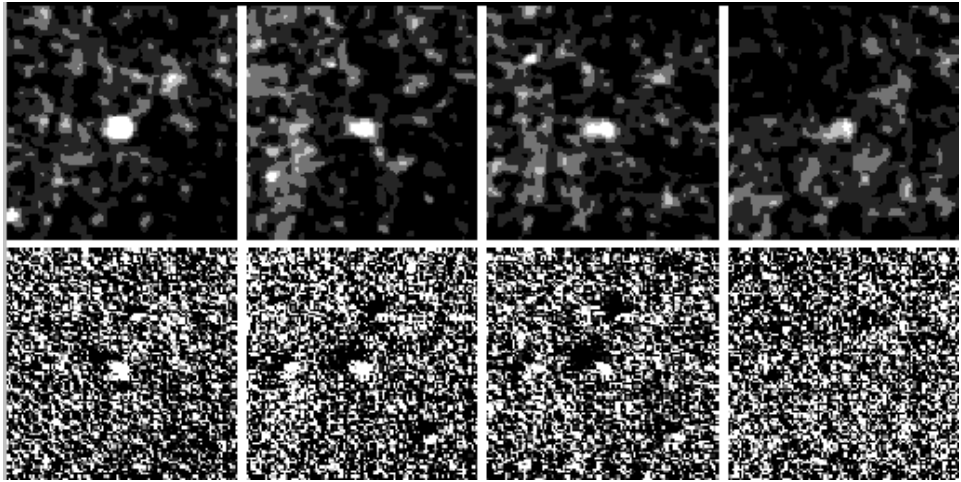,height=2.5in,width=5in}}
\vspace{10pt}
\caption{Difference images of SN1999ay 1,3,4
and 5 months after discovery produced with our new image 
subtraction method (upper panel) and
with the standard one (lower panel). Note that with the new 
method this SN is still detectable after 5 months (upper right)}
\label{figure1}
\end{figure}

Our main scientific goal is to derive from our data the SN rate as a function
of various parameters such as host galaxy type and cluster environment: 
position within the cluster, cluster richness and cluster vs. field. SN 
rates can then be used to determine the current and past star formation 
rates in galaxy clusters \cite{mdp}. Our measured SN rates can replace 
the assumed rates used so far in studies of metal abundances in the 
intracluster gas.
We also intend to study the rate, distribution and properties of 
intergalactic SNe in galaxy clusters. A candidate intergalactic SN we have 
discovered, 
SN 1998fc (see fig. 3), will be discussed below.
Our search is also sensitive to other optical transients, such as AGNs in the
clusters and behind them, flares from tidal disruption of stars by dormant
massive black holes in galactic nuclei and GRB afterglows. We may also detect
the gravitational lensing effect of the clusters on background SNe \cite{kb}.

\section*{First results}

Our program has already discovered 11 spectroscopically 
confirmed SNe at $ z=0.1-0.24,$ (see table 1 and fig. 2) and several 
unconfirmed SNe.    
We have also detected variable stellar objects (some of which are AGN) 
and dozens of asteroids.

\begin{table}[b!]
\caption{Spectroscopically confirmed SNe discovered at the Wise Observatory}
\label{table1}
\begin{tabular}{dddddddc}
   SN& Cluster& Cluster z& Type& SN z& Discovered & R mag & spectrum\\
\tableline
1998cg & Abell 1514 & 0.199 & Ia & 0.12  & 1.5.98 & 18.5 & ESO 3.6m \\
1998eu & Abell 125  & 0.188 & Ia & 0.181 & 14,11.98 & 19.7 & AAT 4m \\
1998fc & Abell 403  & 0.103 & Ia & 0.10  & 20.12.98 & 20.5 & ESO 3.6m \\
1998fd & Field      &       & Ia & 0.24  & 24.12.98 & 21.3 & Keck II 10m \\
1999C  & Abell 914  & 0.195 & Ia & 0.125 & 14.1.99 & 19.6 & Keck II 10m \\
1999ax & Abell 1852 & 0.181 & Ia & 0.09  & 20.3.99 & 18.5 & KPNO 4m \\
1999ay & Abell 1966 & 0.151 & II? & 0.15 & 21.3.99 & 18.0 & KPNO 4m \\
1999cg & Abell 1607 & 0.136 & Ia & 0.14  & 15.4.99 & 19.2 & Keck II 10m \\
1999ch & Abell 2235 & 0.151 & Ia & 0.15  & 13.5.99 & 19.8 & KPNO 4m \\
1999ci & Abell 1984 & 0.124 & Ia & 0.12  & 15.5.99 & 19.3 & KPNO 4m \\
1999ct & Abell 1697 & 0.183 & Ia & 0.18  & 13.6.99 & 20.8 & Keck II 10m \\
\end{tabular}
\end{table}

\begin{figure}[t!] 
\centerline{\epsfig{file=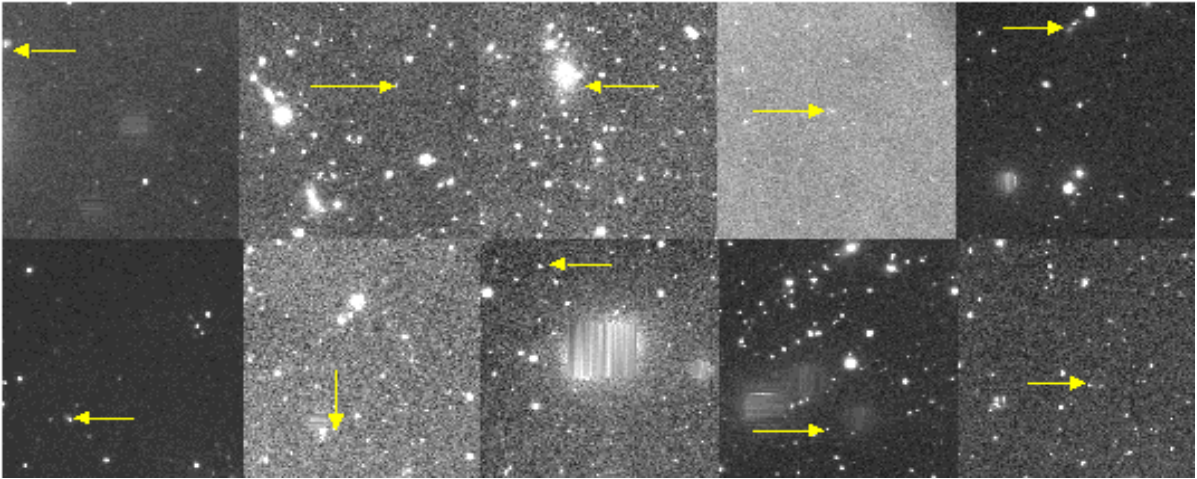,height=2.5in,width=6.25in}}
\vspace{10pt}
\caption{A sample of SN images discovered at the Wise Observatory:
\newline
(From left to right and top to bottom)
SN 1998cg, 1998eu, 1998fc, 1998fd,
1999C, 1999ax, 1999ay, 1999cg, 1999ch, 1999ci}
\label{figure2}
\end{figure}

\section*{Intergalactic SNe and enhanced central metal abundances in clusters}

The existence of a diffuse population of intergalactic stars is supported
by a growing body of observational evidence such as intergalactic planetary nebulae in the Fornax and Virgo clusters \cite{thw} \cite{arn} \cite{ciar} \cite {fre}, 
and intergalactic red giant stars in Virgo \cite{ftv}. Recent imaging of the Coma cluster reveals low surface brightness
emission from a diffuse population of stars \cite{gw}, the origin of which is attributed to galaxy disruption \cite{dmh} \cite {mor}.
Since type Ia SNe are known to occur in all environments, there is no obvious reason 
to assume that such events do not happen within the intergalactic stellar population. SN 1998fc may be such an event. The  
intergalactic stellar population is centrally distributed \cite{dub}.
Therefore, metals produced by intergalactic Ia SNe can provide an elegant
explanation for the central enhancement of metal abundances with type Ia
characteristics, recently detected in galaxy clusters \cite{dw}. 
 
\subsection*{SN 1998fc - An intergalactic SN candidate in Abell 403}

SN 1998fc was detected near the cD galaxy of Abell 403 \cite{gm1}, 
and was
spectroscopically confirmed as a type Ia SN at the cluster redshift
\cite{gm2} \cite{flr}.
The most likely host for this SN, the cD galaxy, is very distant -
at least 78 Kpc away. This may be an intergalactic SN whose
progenitor star was a member of the diffuse intergalactic stellar
population. Alternatively, the host may be a faint dwarf galaxy. The
distribution of "hostless" SNe is expected to be different if the 
progenitors are members of the intergalactic population, centered near the
cluster core \cite{dub}, or members of dwarf galaxies, more abundant in the 
outskirts of galaxy clusters \cite{phi}. Therefore, the nature of such objects
could be resolved with larger number statistics. In any event, the number 
of SNe with undetected hosts relative to the total number of cluster SNe 
can put an upper limit on the intergalactic stellar fraction.

\begin{figure} 
\centerline{\epsfig{file=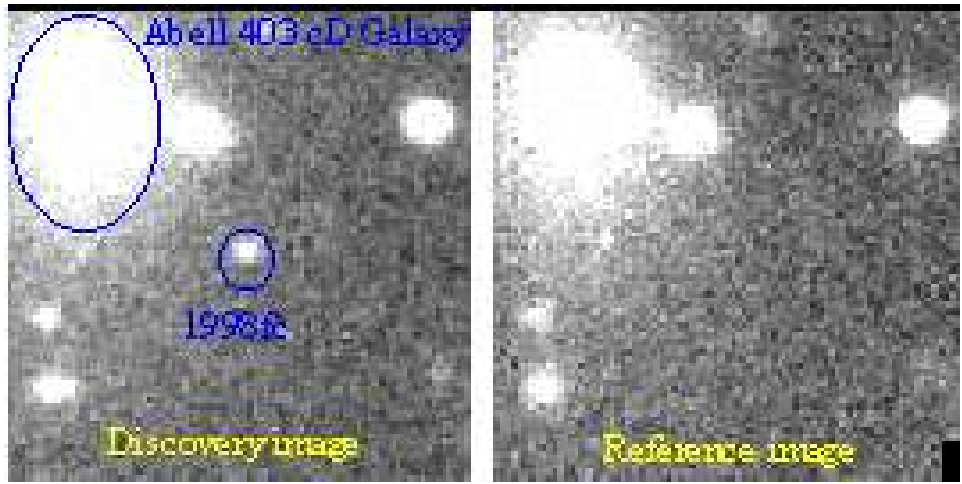,height=2.5in,width=5in}}
\vspace{10pt}
\caption{SN 1998fc - An intergalactic SN candidate in Abell 403.}
\label{figure3}
\end{figure}


\begin{references}
\bibitem{mdp} Madau, P., DellaValle, M., \& Panagia, N. 1998, MNRAS, 297, L17
\bibitem{kb}  Kolatt, T., \& Bartelmann, M. 1998, MNRAS, 296, 763
\bibitem{thw} Theuns, T., \& Warren, S.J. 1996, MNRAS, 284, L11
\bibitem{arn} Arnaboldi, M. et al. 1996, ApJ, 472, 145
\bibitem{ciar} Ciardullo, R. et al. 1998, ApJ, 492, 62
\bibitem{fre} Freeman, K. C. et al. 1999, astro-ph/9910057
\bibitem{ftv} Ferguson, H.C., Tanvir, N.R., \& von Hippel, T. 1998, Nature, 391, 461
\bibitem{gw} Gregg, M.D. \& West, M.J., 1998, Nature, 396, 549 
\bibitem{dmh} Dubinski, J., Mihos, J.C. \& Hernquist, L., 1996, ApJ, 462, 576
\bibitem{mor} Moore, B., et al. 1996, Nature, 379, 613
\bibitem{dub} Dubinski, J., 1999, astro-ph/9902331
\bibitem{dw}  Dupke, R.A. \& White, R.E.III, 1999, ApJ, submitted, astro-ph/9902112
\bibitem{gm1} Gal-Yam, A. \& Maoz, D., 1999, IAUC 7082
\bibitem{gm2} Gal-Yam, A. \& Maoz, D., 1999, IAUC 7093
\bibitem{flr} Filippenko, A.V., et al. 1999, IAUC 7091 
\bibitem{phi} Phillipps, S., Driver, S., Couch, W. J. \& Smith, R. M., 1998, ApJ, 498, 119
\end{references}
\end{document}